\documentclass[danish,a4paper,10pt,
twocolumn,
amsmath,amssymb,floatfix,superscriptaddress, 
aps,
pra,showpacs,longbibliography]{revtex4-1}
\usepackage{graphicx}
\usepackage[english]{babel}
\usepackage{bm}
\usepackage{braket}
\usepackage{float}
\usepackage{mathtools}
\usepackage{dcolumn}
\usepackage[draft]{fixme}
\usepackage{color}
\usepackage{lipsum}
\definecolor{red}{rgb}{1,0,0}
\definecolor{blue}{rgb}{0,0,1}

\begin{document}

\title{Analytic modeling of structures in attosecond transient-absorption spectra}
\author{J\o rgen~Johansen~R\o rstad}
\affiliation{Department of Physics and Astronomy, Aarhus University, 8000 Aarhus C, Denmark}
\author{Jens~E.~B\ae kh\o j}
\affiliation{Department of Physics and Astronomy, Aarhus University, 8000 Aarhus C, Denmark}
\affiliation{Department of Physics and Astronomy, Louisiana State University, Baton Rouge, Louisiana 70803, USA}
\author{Lars~Bojer~Madsen}
\affiliation{Department of Physics and Astronomy, Aarhus University, 8000 Aarhus C, Denmark}

\date \today
\begin{abstract}
Attosecond transient-absorption spectroscopy (ATAS) is an established method for exploring electron dynamics on the subfemtosecond time scale. ATAS spectra contain certain ubiquitous features, such as oscillating fringes, light-induced structures, and hyperbolic sidebands, representing physical processes. We derive closed analytical expressions describing these features, based on a three-level system responding adiabatically to the influence of an infrared field in conjunction with an extreme ultraviolet pulse, and use He to illustrate the theory. The validity of the formulas is substantiated by comparing their predictions with spectra calculated numerically by the time-dependent Schr\"odinger equation. The closed analytical forms and the details of their derivation resolve the origins of the features.
\end{abstract}

\maketitle

\section{Introduction}

Since its inception, attosecond transient absorption spectroscopy (ATAS) has served as a useful tool in the exploration of electron dynamics in atoms and molecules \cite{goulielmakis2010,Beck2015,wu2016}. This fully optical method makes use of a femtosecond infrared (IR) pulse, which dresses the system, acting together with an attosecond extreme ultraviolet (XUV) pulse whose spectrum is modified during the interaction. Information can be extracted from this modified spectrum and gathered in rich spectrograms demonstrating absorption and emission of light, where time-resolution has been indirectly introduced by the variable delay between the two pulses. ATAS has enabled the exploration of various processes on their natural time scales, as exemplified by numerous investigations; for instance, in the observation of strong field dynamics and ionization in several systems \cite{kobayashi2017,wirth2011,pabst2012,sabbar2017}, autoionization in Ar \cite{wang2010}, and the control of wave packets in He \cite{ott2014}. The method has also been successfully applied to molecular systems, such as in experimental studies on $\text{N}_2$  \cite{warrick2016,warrick2017} and in theoretical studies on charge migration \cite{hollstein2017} and the effects of nuclear dynamics \cite{baekhoej2015}.

\begin{figure}
\includegraphics[width=0.48\textwidth]{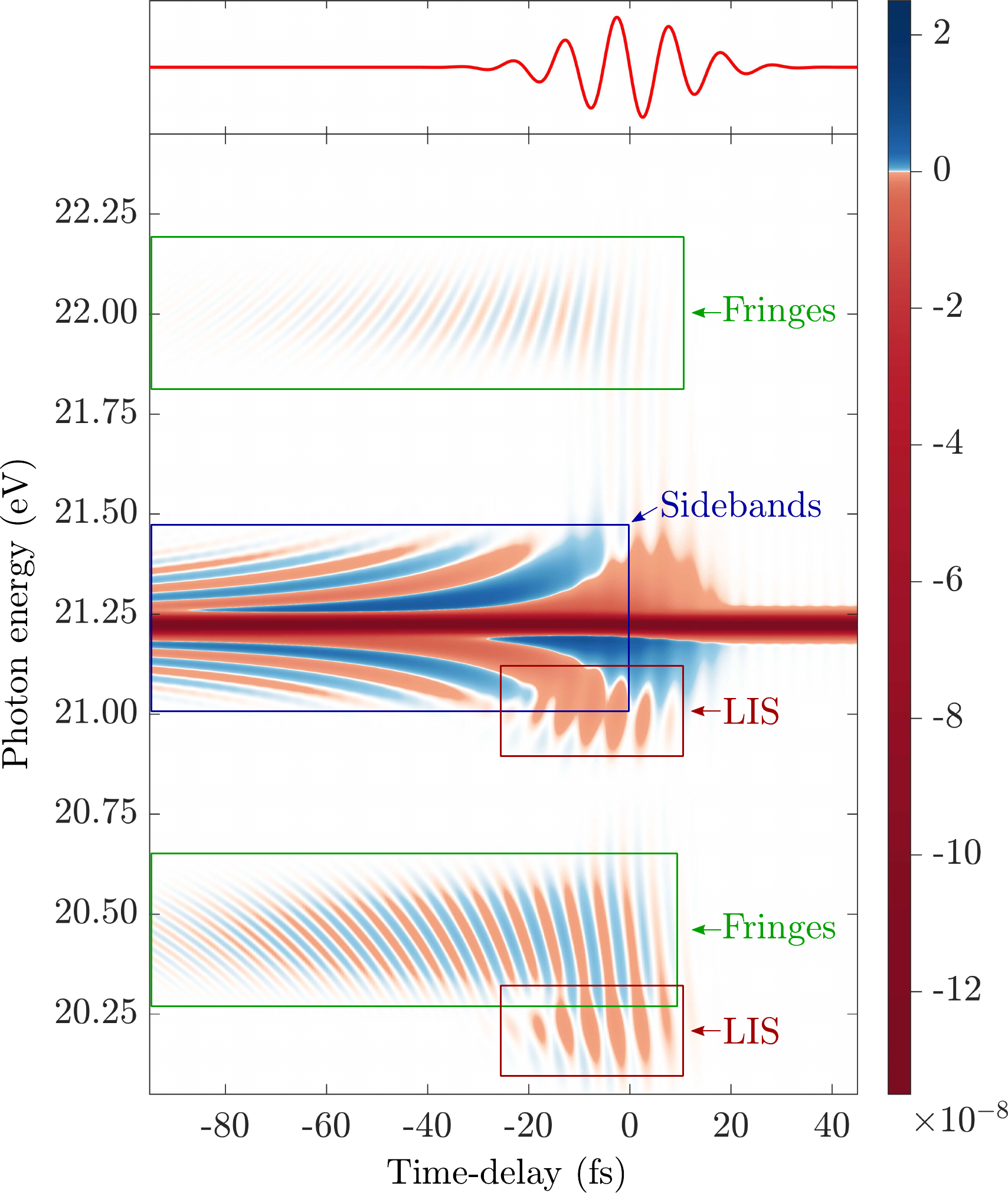}
\caption{\label{fig:highlights} Attosecond transient absorption spectrum $\tilde{S}(\omega,\tau)$ [Eq.~\eqref{theory:A:response_1}] based on the adiabatic three-level model developed in Sec.~\ref{subsection:3level}, exemplified by He. Pertinent features are highlighted: oscillating fringes centered around $E(1s2p)\pm2\omega_\text{IR}$, i.e., at $22.00$ and $20.45$ eV; hyperbolic sidebands centered around $E(1s2p)=21.22$ eV; and light-induced structures (LISs) centered around $E(1s2s)\pm\omega_\text{IR}$, i.e., at $21.01$ and $20.23$ eV. The top panel illustrates the infrared pulse centered at $\tau=0$ fs. The color scale to the right shows the signal strength in arbitrary units. Analytical expressions will be derived for the highlighted features. For pulse parameters, see the text following Eq.~\eqref{theory:A:pulse}.}
\end{figure}

For atoms, the features found in ATA spectrograms correspond to physical processes undergone by electrons, and various descriptions of these processes have been offered. Examples include the presence of oscillating fringes signifying different forms of interference \cite{chen2013,chini2014}, light-induced structures (LISs) signaling the impact of virtual intermediate states in two-photon absorption processes \cite{chen2012}, hyperbolic sidebands (SBs) associated with perturbed free-induction decay \cite{lindberg1988,britocruz1988}, and Autler-Townes splitting of the absorption lines indicative of resonant population transfer between states \cite{autler1955,pfeiffer2012,wu2013}. The features of which the present study is concerned are highlighted in the ATA spectrogram seen in Fig.~\ref{fig:highlights}, which is calculated using the method described in Sec.~\ref{subsection:3level}. Theoretical investigations of these features generally rely on either directly solving the time-dependent Schr\"odinger equation (TDSE), possibly in a finite basis of field-free bound states \cite{ott2014}, or by devising models capable of reproducing some desired characteristics \cite{chen2013lip}. 

We derive analytical closed forms representing the oscillating fringes, LISs and SBs adjacent to the main absorption lines (see Fig.~\ref{fig:highlights}). These expressions enable a clearer understanding of the origin of the features and their underlying physical processes. Moreover, there is great interest in extracting attosecond dynamics directly from experimental ATA spectra, which requires simple yet precise models. The basis of this derivation is a three-level system where the two excited levels are expressed in terms of adiabatic states, an essential premise for which is that the IR pulse changes slowly relative to the bound electron motion, hence fulfilling the adiabatic condition. A three-level system subject to the adiabatic condition has previously been considered numerically~\cite{dong2015}, but an analytical treatment has been lacking. Given appropriate parameters for the system and incident electric fields, this three-level setup can serve as a good approximation to certain physical systems, and is simple enough to allow for an analytical analysis. We choose a setup where only one of the excited states, designated as the bright state, can be accessed through a dipole allowed transition from the ground state. For the other excited state, designated as the dark state, there exists no dipole allowed coupling with the ground state. These excited energy levels lie within the broad spectrum of frequencies in the XUV pulse. The IR pulse consists of low-frequency components and is too weak to excite the system via multiphoton absorption. The IR pulse will, however, induce a mixing of the excited states. This mixture of field-free states serves as a basis through which the adiabatic states can be expressed. We exemplify our model by the three lowest levels of He, where the bright state is $\ket{1s2p}$ and the dark state is $\ket{1s2s}$.

The paper is organized as follows. Section~\ref{section:theory} contains the theoretical framework of all relevant methods and models. Specifically, Sec.~\ref{subsection:response} introduces the single-atom response function, which describes the modification of the XUV spectrum in ATAS; Sec.~\ref{subsection:TDSE} summarizes the method based on the TDSE and compares a twenty-level system with a three-level system in order to support the validity of the three-level model; and Sec.~\ref{subsection:3level} contains the description of the analytic modeling, where certain calculations related to the Fourier transform of the time-dependent dipole moment are relegated to the Appendix. In Sec.~\ref{section:results} the comparison between the various methods and the analytical expressions is shown, followed by individual scrutiny of the three features of interest: SBs, LISs and fringes. Finally, Sec.~\ref{sec:conclusion} concludes and offers an outlook. Atomic units ($\hbar=e=m_e=a_0=1$) are used throughout, unless otherwise indicated.

\section{\label{section:theory}Theory and discussion} 
This section starts by introducing the response function of the atomic system, from which we acquire the delay-dependent ATA spectrum. The calculation of this response function for a system of bound states by a method involving the TDSE is summarized, and closed analytic forms corresponding to specific features of the ATA spectrum are derived. 

\subsection{\label{subsection:response}Single-atom response function} 
The moderate intensities of the fields used in this study enable the application of certain weak-field approximations. However, if we consider the density of photons, the intensities are sufficiently high that a classical treatment of the field is justified. In such a semi-classical approach only the atomic system is given a full quantum mechanical description, significantly reducing the complexity of the calculations. An attosecond XUV pulse excites the system at a given time $t=\tau$ and a femtosecond IR pulse induces dynamics in the excited states. We fix the pulse center of the IR pulse at $t=0$, so that the delay $\tau$ represents the relative position of the XUV pulse center. Both fields are linearly polarized in the $z$ direction.

The delay-dependent single-atom response describes the modulation of the XUV field interacting with an IR field-dressed atom. Realistically a full description must include the medium of propagation, but for sufficiently dilute gases the single-atom response is a valid approximation~\cite{gaarde2011}. This response function is given directly (see Ref.~\cite{baggesen2012} for details):
\begin{equation}
\tilde{S}(\omega,\tau)=\frac{4\pi n\omega}{c}\text{Im}[\tilde{\mathcal{E}}^*_{\text{in}}(\omega,\tau)\tilde{d}(\omega,\tau)].\label{theory:A:response_1}
\end{equation}
Here $n$ is the density of atoms in the sample, $c\simeq137$ is the speed of light, $\tilde{\mathcal{E}}_\text{in}(\omega,\tau)$ is the incoming XUV field in the frequency domain, and $\tilde{d}(\omega,\tau)$ is the expectation value of the dipole moment's $z$ component in the frequency domain. The tilde is used to denote Fourier-transformed quantities, where we use the convention $\tilde{f}(\omega)=\frac{1}{\sqrt{2\pi}}\int_{-\infty}^\infty dt\, f(t) e^{-i\omega t} $. In Eq.~\eqref{theory:A:response_1}, a negative value corresponds to absorption, and a positive value corresponds to emission of light. There exists in the literature several expressions equivalent to the response function in Eq.~\eqref{theory:A:response_1}. For an overview and discussion of these, see Ref.~\cite{baekhoej2015}.

All fields are obtained from $\mathcal{E}(t)=-\partial_tA(t)$, where $A(t)$ is the vector potential given by
\begin{equation}
A(t)=A_0\exp\left[-\frac{(t-t_c)^2}{T^2/4}\right]\cos{[\omega(t-t_c)]}\label{theory:A:pulse},
\end{equation}
with $A_0=\mathcal{E}_0/\omega$ and $\omega$ being the angular frequency. $\mathcal{E}_0$ is related to the intensity by $I\propto|\mathcal{E}_0|^2$. In Eq.~\eqref{theory:A:pulse}, $t_c$ represents the center of the pulse, and the period is $T=N_cT_c=N_c\frac{2\pi}{\omega}$, where $N_c$ is the number of cycles in the pulse and $T_c$ is the period of one cycle. The period $T$ is connected with the temporal full width at half maximum (FWHM) of the field through $T_\text{FWHM}=\sqrt{\log{2}}T$. For the IR pulse $t_c=0$, and for the XUV pulse $t_c=\tau$. The specific pulse parameters used herein are as follows: $\lambda_\text{IR}=3200$~nm, $\lambda_\text{XUV}=50$~nm, $I_\text{IR}=6\times10^{10}$ W/cm${}^2$, $I_\text{XUV}=5\times10^7$ W/cm${}^2$, $T_\text{IR}=32.02$~fs ($N_{c,\text{IR}}=3$), and $T_\text{XUV}=330$~as ($N_{c,\text{XUV}}=2$). We arbitrarily set the density of atoms $n=1$ [see Eq.~\eqref{theory:A:response_1}].

\subsection{\label{subsection:TDSE}Time-dependent Schr\"odinger equation}

In order to obtain the response of the system, we calculate the expectation value of the time-dependent dipole moment $\braket{d(t)}\equiv\braket{\Psi(t)|d|\Psi(t)}$, with $d = -\sum_k z_k$ being the dipole moment operator in the polarization direction. We express the quantum state of the system as 
\begin{equation}
\ket{\Psi(t)}=\sum_{n=0}^N c_n(t)e^{-iE_nt}\ket{\phi_n}, \label{theory:B:delta}
\end{equation}
with $\ket{\phi_n}$ being an energy eigenstate of the unperturbed system corresponding to energy $E_n$ and $N$ being the number of bound states in a finite basis of field-free states \cite{chini2014}. The full Hamiltonian of the system is $H=H_0+H'(t)$, where the time-dependent part accounts for the electric fields. Within the dipole approximation and in the length gauge it is given as
\begin{equation}
H'(t)=-\mathcal{E}(t)d,
\end{equation} 
with $\mathcal{E}(t)$ being the total field, including both the XUV and IR field. By inserting Eq.~\eqref{theory:B:delta} into the TDSE and projecting on $\bra{\phi_m}$, we obtain a set of coupled differential equations for the coefficients $c_m(t)$. These equations can be solved numerically, and the time-dependent dipole moment is determined by
\begin{equation}
\braket{\Psi(t)|d|\Psi(t)}=\sum\limits_{m,n=0}^Nc_m^*(t)c_n(t)e^{-i(E_n-E_m)t}d_{mn},\label{theory:B:d_TDSE}
\end{equation}
with $d_{mn}=\braket{\phi_m|d|\phi_n}$. We use He to illustrate our theory, with energies and transition dipole moments given in Ref.~\cite{drake2006}.

In experiments, the time-dependent dipole moment is subject to dephasing due to effects such as collisional broadening and finite detector resolution. These effects can be replicated by imposing a window function $W(t-\tau)$ which dampens $\braket{d(t)}$ over some appropriate time $T_0$, starting at the moment of incidence for the XUV pulse \cite{baekhoej2015,wu2016}. We define the window function as
\begin{equation}
W(t-\tau)=
\begin{dcases}
1 & (t<\tau)\\
\exp{\Big[-\frac{(t-\tau)^2}{2T_0^2} \Big]}	& (\tau \leq t),
\end{dcases} \label{theory:B:window}
\end{equation}
where $2\sqrt{2\ln(2)}T_0=T_\text{FWHM}$ is the FWHM of the corresponding Gaussian. $T_\text{FWHM}$ is chosen to be large enough that the features of interest in the ATA spectrum remain unaffected as the window function is imposed, which for the parameters in this work is $T_\text{FWHM}/2=100$ fs. After multiplying Eq.~\eqref{theory:B:d_TDSE} by the window function from Eq.~\eqref{theory:B:window} and taking the Fourier transform returns $\tilde{d}(\omega,\tau)$, the response function can be determined from Eq.~\eqref{theory:A:response_1}.

\begin{figure}
\includegraphics[width=0.48\textwidth]{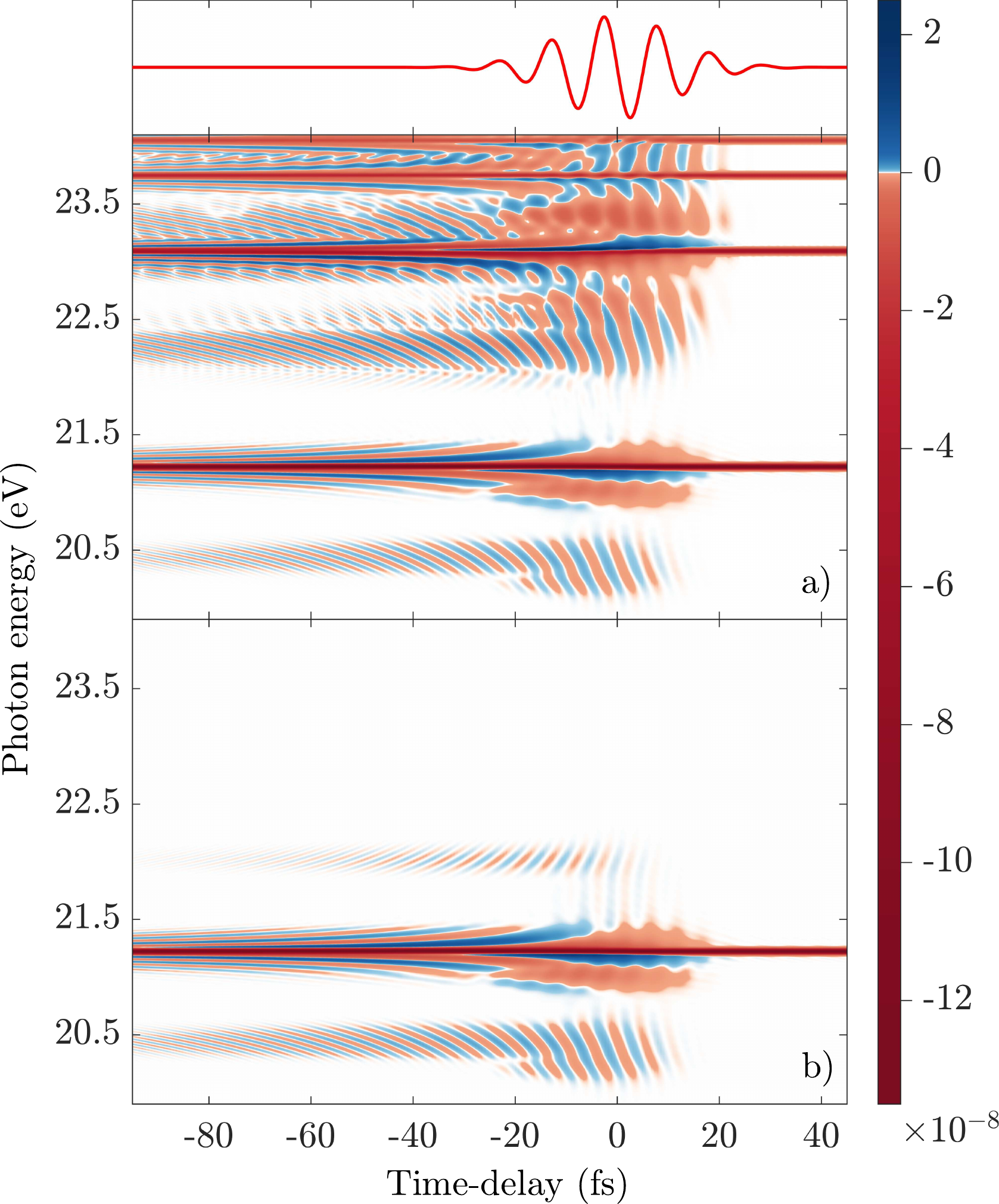}
\caption{\label{fig:TDSE20vsTDSE3} Attosecond transient absorption spectra $\tilde{S}(\omega,\tau)$ of He calculated by Eq.~\eqref{theory:A:response_1}, where $\tilde{d}(\omega,\tau)$ is determined via the TDSE [see Eq.~\eqref{theory:B:d_TDSE}]. In panel (a), the results are obtained with a basis consisting of the $N+1=20$ lowest levels of He. In panel (b), the results are obtained with a basis consisting of only the $N+1=3$ lowest levels of He: the ground state, fixed at $0$ eV, $\ket{1s2s}$ at $20.62$ eV, and $\ket{1s2p}$ at $21.22$ eV. The top panel illustrates the infrared pulse centered at $\tau=0$ fs. The color scale to the right shows the signal strength in arbitrary units. For pulse parameters, see the text following Eq.~\eqref{theory:A:pulse}.}
\end{figure} 

The method described in this section is meant to provide a reference to which we can compare our analytic model, which consists of three energy levels. For this to be a valid comparison, we also require the TDSE calculations to be accurate for a three-level system [Eq.~\eqref{theory:B:delta} with $N+1=3$]. We achieve this by comparing solutions based on $N+1=20$ and $N+1=3$ bound states of the present model system, He, and verifying that they agree in the relevant range of energies.  A comparison of the two cases is shown in Fig.~\ref{fig:TDSE20vsTDSE3}, where we have used the pulse parameters given in the text following Eq.~\eqref{theory:A:pulse} and the window function from Eq.~\eqref{theory:B:window}. The energy range for which we require agreement between the two solutions extends from $\sim20$ to $\sim22.4$ eV. The only disparity is seen around $22.3$ eV, where there is a change as we go from twenty levels in Fig.~\ref{fig:TDSE20vsTDSE3} (a) to three levels in Fig.~\ref{fig:TDSE20vsTDSE3} (b). In this region, oscillating fringes corresponding to the $\ket{1s3p}$ state are overlapping with the equivalent feature corresponding to the $\ket{1s2p}$ state. As the $\ket{1s3p}$ state is removed from the calculations, its corresponding fringes vanish and uncover the fringes associated with the $\ket{1s2p}$ state. Consequently, the three-level TDSE calculations can be considered accurate in the energy range of interest.

\subsection{\label{subsection:3level}Adiabatic three-level model}
In this section we obtain analytical expressions corresponding to certain features in the ATA spectrum of a system with three energy levels, the first step of which is to determine the time-dependent dipole moment of the system, $\braket{\Psi(t)|d|\Psi(t)}$. We then apply perturbation theory and a number of approximations, sequentially reducing it to a form which is sufficiently simple to allow for an analytic Fourier transform. The response function is finally obtained from Eq.~\eqref{theory:A:response_1}.

For systems evolving adiabatically or near adiabatically, an expansion in the adiabatic states $\ket{\phi_{na}(t)}$, defined through
\begin{equation}\label{theory:C:adiabatic_equation}
H(t)\ket{\phi_{na}(t)}=E_{na}(t)\ket{\phi_{na}(t)}, 
\end{equation} is often preferable. Equation~\eqref{theory:C:adiabatic_equation} expresses that the adiabatic states $\ket{\phi_{na}(t)}$ are the instantaneous eigenstates of the Hamiltonian of our system, with the eigenenergies $E_{na}(t)$. We may express the solution to the TDSE in the adiabatic basis as
\begin{equation}
\begin{aligned}
\ket{\Phi(t)}&=b_0(t)\ket{\phi_0}+a_1(t)\ket{\phi_{1a}(t)}e^{-i\int_{\tau}^tdt'E_{1a}(t')}\\
&\quad+a_2(t)\ket{\phi_{2a}(t)}e^{-i\int_{\tau}^tdt'E_{2a}(t')}, \label{theory:C:3level}
\end{aligned}
\end{equation}
with the ground-state energy $E_0$ set to zero and $a_1(t)$, $a_2(t)$ being the coefficients of the adiabatic states.  Since the intensity of the XUV field used in our calculation is weak, $a_1(t)$ and $a_2(t)$ will be small at all times. In Eq.~\eqref{theory:C:3level}, we have made the assumption that the ground state does not change appreciably under the influence of the fields, and denoted its coefficient by $b_0(t)$. $e^{-i\int_{\tau}^tdt'E_{na}(t')}$ ($n=1,2$) is the phase factor for the adiabatic states with time-dependent energies.

To obtain the adiabatic states and associated energies, we start by considering the full Hamiltonian of our system
\begin{equation}
\underline{\underline{H_f}}=
\begin{bmatrix}
0 & d_{01}\mathcal{E}_{\text{XUV}}(t) & 0\\
d_{01}\mathcal{E}_{\text{XUV}}(t) & E_{1} & d_{12}\mathcal{E}_{\text{IR}}(t) \\
0 & d_{12}\mathcal{E}_{\text{IR}}(t) & E_{2}
\end{bmatrix},
\end{equation}
which illustrates the coupling, due to the XUV pulse, between the ground state and the first excited state; and similarly for the two excited states, coupled by the IR pulse. The couplings are chosen such that $\ket{\phi_1}$ is the bright state and $\ket{\phi_2}$ is the dark state with respect to a dipole transition from the field-free ground state $\ket{\phi_0}$. 

We can reduce the system by noting that after the XUV pulse has populated the first excited state, the ground state is approximately constant and we effectively have a two-level system specified by 
\begin{equation}
\underline{\underline{H}}=
\begin{bmatrix}
E_{1} & d_{12}\mathcal{E}_{\text{IR}}(t) \\
d_{12}\mathcal{E}_{\text{IR}}(t) & E_2
\end{bmatrix}.
\end{equation}
The instantaneous eigenvalues of this Hamiltonian are
\begin{align}
E_{1a}(t) &\approx E_1+\frac{d_{12}^2\mathcal{E}^2_\text{IR}(t)}{E_1-E_2}\label{theory:C:eig1},\\
E_{2a}(t) &\approx E_2-\frac{d_{12}^2\mathcal{E}^2_\text{IR}(t)}{E_1-E_2}\label{theory:C:eig2},
\end{align}
where we have performed a Taylor expansion about $\mathcal{E}_\text{IR}(t)=0$ and neglected terms of third order or higher in the IR field. The normalized adiabatic eigenstates are
\begin{equation}
\begin{aligned}
\ket{\phi_{1a}(t)}&=\frac{d_{12}\mathcal{E}_\text{IR}(t)}{\sqrt{d^2_{12}\mathcal{E}^2_\text{IR}(t)+[E_{1a}(t)-E_1]^2}}\ket{\phi_1}\label{theory:C:eigf1}\\
&\quad+\frac{E_{1a}(t)-E_1}{\sqrt{d^2_{12}\mathcal{E}^2_\text{IR}(t)+[E_{1a}(t)-E_1]^2}}\ket{\phi_2}
\end{aligned}
\end{equation}
and
\begin{equation}
\begin{aligned}
\ket{\phi_{2a}(t)}&=\frac{E_{2a}(t)-E_2}{\sqrt{d^2_{12}\mathcal{E}^2_\text{IR}(t)+[E_{2a}(t)-E_2]^2}}\ket{\phi_1}\label{theory:C:eigf2}\\
&\quad+\frac{d_{12}\mathcal{E}_\text{IR}(t)}{\sqrt{d^2_{12}\mathcal{E}^2_\text{IR}(t)+[E_{2a}(t)-E_2]^2}}\ket{\phi_2},
\end{aligned}
\end{equation}
which appropriately reduce to the field-free states in the limit $\mathcal{E}_\text{IR}(t)\to~0$. 

To determine the coefficients $a_1(t)$ and $a_2(t)$ of Eq.~\eqref{theory:C:3level}, it is convenient to first express the same quantum state in terms of the field-free basis states: 
\begin{equation}
\ket{\Phi(t)}=b_0(t)\ket{\phi_0}+b_1(t)\ket{\phi_1}+b_2(t)\ket{\phi_2}, \label{theory:C:ffstate}
\end{equation}
where the coefficients are
\begin{equation}
\begin{aligned}
b_1(t) &= a_1(t)e^{-i\int_{\tau}^t dt' E_{1a}(t')}\braket{\phi_1|\phi_{1a}(t)}\\
&\quad+a_2(t)e^{-i\int_{\tau}^t dt'E_{2a}(t')}\braket{\phi_1|\phi_{2a}(t)},
\end{aligned}
\end{equation}
and
\begin{equation}
\begin{aligned}
b_2(t) &=a_1(t)e^{-i\int_{\tau}^t \text{dt}' E_{1a}(t')}\braket{\phi_2|\phi_{1a}(t)}\\
&\quad+a_2(t)e^{-i\int_{\tau}^t \text{dt}' E_{2a}(t')}\braket{\phi_2|\phi_{2a}(t)}.
\end{aligned}
\end{equation}
The overlaps $\braket{\phi_m|\phi_{na}(t)} (n,m=1,2)$  can be obtained from Eqs.~\eqref{theory:C:eigf1} and \eqref{theory:C:eigf2}:
\begin{align}
\braket{\phi_1|\phi_{1a}(t)}&=\braket{\phi_2|\phi_{2a}(t)}\approx1-\frac{d_{12}^2\mathcal{E}_\text{IR}^2(t)}{2(E_1-E_2)^2},\label{theory:C:overlap1} \\
\braket{\phi_1|\phi_{2a}(t)}&=-\braket{\phi_2|\phi_{1a}(t)}\approx-\frac{d_{12}\mathcal{E}_\text{IR}(t)}{E_1-E_2},\label{theory:C:overlap2}
\end{align}
where we have neglected terms of third order or higher in the IR field after an expansion about $\mathcal{E}_\text{IR}(t)=0$.

Inserting Eq.~\eqref{theory:C:ffstate} into the TDSE and projecting on to each of the excited field-free states $\ket{\phi_1}$ and $\ket{\phi_2}$, we obtain equations for the coefficients $\dot{a}_n(t)$ (dot denotes time derivative), which we then treat perturbatively. If the system remains unperturbed, we expect it to stay in the ground-state indefinitely; $b^{(0)}_0(t)=1$, $a^{(0)}_1(t)=a^{(0)}_2(t)=0$ (superscripts denote the order of approximation). To obtain the first order approximations for the coefficients, we insert the zeroth-order coefficients back into our differential equations. Neglecting third-order or higher terms of IR field in the pre-exponential factor, the equations become
\begin{align}
	\dot{a}_1^{(1)}(t)&\approx id_{01}\mathcal{E}_\text{XUV}(t)e^{i\int_{\tau}^tdt'E_{1a}(t')}\left[ 1-\frac{d_{12}^2\mathcal{E}_\text{IR}^2(t)}{2(E_1-E_2)^2}\right]\label{theory:C:adot1}, \\
	\dot{a}_2^{(1)}(t)&\approx -\frac{id_{01}d_{12}}{E_1-E_2}\mathcal{E}_\text{XUV}(t)\mathcal{E}_{\text{IR}}(t)e^{i\int_{\tau}^tdt'E_{2a}(t')}\label{theory:C:adot2}.
\end{align}
The change in the coefficient of the ground state is assumed to be negligible throughout the interaction, and so we have set $b_0(t)=1$. 

From this point on, it is practical to work directly with the expectation value of the time-dependent dipole moment. To leading order in the small coefficients $b_1(t)$ and $b_2(t)$, we obtain
\begin{equation}
\braket{\Phi(t)|d|\Phi(t)}=\braket{d(t)}=d_{01}\left[b_1(t)+b_1^*(t)\right],
\end{equation}
where we have used that $d_{02}=0$. This can be simplified further by observing that the $b_1(t)$ coefficient contains factors $e^{-iE_1t}$ and $e^{-iE_2t}$, which upon Fourier transform correspond to shifts to negative frequencies. We need only consider positive frequencies, and consequently only $b_1^*(t)$ is kept, resulting in
\begin{equation}
\begin{aligned}
	\braket{d(t)}&=d_{01}\bigg[a_1^*(t)e^{i\int_{\tau}^t dt' E_{1a}(t')}\braket{\phi_1|\phi_{1a}(t)}^* \\
&\quad+a_2^*(t)e^{i\int_{\tau}^t dt'E_{2a}(t')}\braket{\phi_1|\phi_{2a}(t)}^*\bigg],
\end{aligned} \label{theory:C:dipmom}
\end{equation}
which must be multiplied by the window function $W(t-\tau)$ from Eq.~\eqref{theory:B:window} to account for dephasing. 

We now have a complete model of the three-level system. The numerical solutions of Eqs.~\eqref{theory:C:adot1} and \eqref{theory:C:adot2} can be inserted into Eq.~\eqref{theory:C:dipmom} to obtain the time-dependent dipole moment. Numerically calculating the Fourier transform and inserting into Eq.~\eqref{theory:A:response_1}, along with the Fourier transform of the XUV pulse, yields the response function of the system. 

A series of approximations are required to progress further toward an analytical form of the response function. First, we assume that the XUV pulse is sufficiently brief that its time-dependence can be approximated by the Dirac $\delta$ function \cite{wu2016}
\begin{equation} \label{theory:C:delta}
\mathcal{E}_\text{XUV}(t)\to \alpha \delta(t-\tau),
\end{equation}
where $\alpha$ is constant in a given set of field parameters and is determined by comparison with an equivalent numerical solution for the response function. In the present calculations, $\alpha=1.61\times10^{-4}$. The replacement in Eq.~\eqref{theory:C:delta} trivializes the task of solving Eqs.~\eqref{theory:C:adot1} and \eqref{theory:C:adot2}:
\begin{align}
	a_1^{(1)}(t)&=\int_\tau^tdt' \dot{a}_1^{(1)}(t')\nonumber\\
	&=i\alpha d_{01}\theta(t-\tau)\left[1-\frac{d_{12}^2\mathcal{E}_\text{IR}^2(\tau)}{2(E_1-E_2)^2} \right]\label{theory:C:a1_delta},\\
	a_2^{(1)}(t)&=\int_\tau^tdt' \dot{a}_2^{(1)}(t')=-\frac{i\alpha d_{01}d_{12}}	{E_1-E_2}\theta(t-\tau)\mathcal{E}_\text{IR}(\tau), \label{theory:C:a2_delta}
\end{align}
where $\theta(t-\tau)$ is the Heaviside step function. As will be shown later, the two terms in Eq.~\eqref{theory:C:dipmom} correspond to distinct features in the ATA spectrum. The first gives rise to delay-dependent SBs \citep{chini2014,chen2013} and oscillating fringes \cite{chini2014} associated with the $E_1$ energy level; the second generates LISs \cite{chen2012} associated with the $E_2$ energy level (see Fig.~\ref{fig:TDSE3vsFULL}). We treat the terms separately. 

\subsubsection{\label{subsubsection:sidebands}Closed analytical form of SBs and oscillating fringes and discussion of physical origins} 
From the first term in Eq.~\eqref{theory:C:dipmom}, we have [see Eq.~\eqref{theory:C:a1_delta}]
\begin{equation}
\begin{aligned}
\braket{d(t)}_1=&-i\alpha d_{01}^2\theta(t-\tau)\left[1-\frac{d_{12}^2\mathcal{E}_\text{IR}^2(\tau)}{2(E_1-E_2)^2} \right]\\
&\times e^{i\int_{\tau}^t dt' E_{1a}(t')}\braket{\phi_1|\phi_{1a}(t)}^*. \label{theory:C:d_SB1}
\end{aligned}
\end{equation} From Eq.~\eqref{theory:C:eig1}, we see that the exponential factor in Eq.~\eqref{theory:C:d_SB1} can be separated as \
\begin{equation} \label{theory:C:sep}
\begin{aligned}
&\exp{[i\int_\tau^t dt' E_{1a}(t')]}\\
&\,\,\,=\exp{[iE_1(t-\tau)]}\exp{[id_{12}^2(E_1-E_2)^{-1}\int_\tau^t dt'\mathcal{E}_\text{IR}^2(t')   ]}.
\end{aligned}
\end{equation} 
We expand the exponential factor in Eq.~\eqref{theory:C:sep} to first order, which is accurate because $\mathcal{E}_\text{IR}^2(t)$ is small:
\begin{equation}
e^{\frac{id_{12}^2}{E_1-E_2}\int_\tau^t dt'\mathcal{E}_\text{IR}^2(t')   }\approx1+\frac{id_{12}^2}{E_1-E_2}\int_\tau^t dt'\mathcal{E}_\text{IR}^2(t'). \label{theory:C:expansion}
\end{equation}
We then obtain [see Eq.~\eqref{theory:C:overlap1}]
\begin{equation} \label{theory:C:d_SB2}
\begin{aligned}
\braket{d(t)}_1=&-i\alpha d_{01}^2\theta(t-\tau)e^{iE_1(t-\tau)}\\
&\times\bigg\{1-\frac{d_{12}^2\left[\mathcal{E}_\text{IR}^2(t)+ \mathcal{E}_\text{IR}^2(\tau)\right]}{2(E_1-E_2)^2}\\
&+\frac{id_{12}^2}{E_1-E_2}\int_\tau^t dt' \mathcal{E}_\text{IR}^2(t')+O(\mathcal{E}_{0,\text{IR}}^4)\bigg\},
\end{aligned}
\end{equation}
where the terms of fourth order or higher in the IR field are neglected. 

We are interested in the features of the ATA spectrum that arise from temporary changes in the time-dependent dipole moment due to the influence of the IR field on the excited states. The Heaviside function $\theta(t-\tau)$ and the window function $W(t-\tau)$ impose restrictions on infinite oscillations that otherwise would correspond to discrete values in the frequency domain. We proceed by setting $\theta(t-\tau)=1$, excluding $W(t-\tau)$, and dropping the two terms in Eq.~\eqref{theory:C:d_SB2} whose only time dependence is in the factor $e^{iE_1t}$:
\begin{equation} \label{theory:C:d_SB3}
\begin{aligned}
\braket{d(t)}_1&=\frac{\alpha d_{01}^2d_{12}^2}{E_1-E_2}e^{-iE_1\tau}e^{iE_1t}\\
&\quad\times\left[\int_\tau^tdt'\mathcal{E}_\text{IR}^2(t')+\frac{i\mathcal{E}_\text{IR}^2(t)}{2(E_1-E_2)}\right].
\end{aligned}
\end{equation}

Neglecting $W(t-\tau)$ and $\theta(t-\tau)$ implies that there will be a nonzero population in the excited states at all times. Before this approximation, the IR pulse could not influence the system prior to the arrival of the XUV pulse, so beyond relatively small positive delays $\tau$ the ATA spectrum would simply exhibit a single-photon absorption pattern, whereas for relatively large negative $\tau$ the system would have time to dephase prior to the arrival of the IR pulse (see Fig.~\ref{fig:TDSE20vsTDSE3}). After this approximation, the ATA spectrum is symmetric about $\tau=0$ and does not weaken at large negative $\tau$. The influence of the IR field at times $t<\tau$ enabled by setting $\theta(t-\tau)=1$ does not have an effect at large negative delay, as the IR field will be zero for $t<\tau$. In the intermediate region where $\tau$ approaches zero, the trailing part of the dipole moment (where $t>\tau$) dominates the leading part (where $t<\tau$), and the results will be similar to the case where $\theta(t-\tau)$ is included, as in the numerical calculations. As $\tau$ approaches zero, the difference between the trailing and leading parts of the dipole moment decreases  and vanishes at $\tau=0$. Finally, dropping of the term that only depends on time in the factor $e^{iE_1t}$ causes the main absorption line at $E=E_1$, corresponding to single photon absorption, to vanish. This poses no problem, as we set out to explain analytically only the features highlighted in Fig.~\ref{fig:highlights}.

The Fourier-transformed time-dependent dipole moment $\tilde{d}_1(\omega,\tau)$ can now be obtained from Eq.~\eqref{theory:C:d_SB3} (see the Appendix for details). The Fourier transform of the XUV field is simply $\tilde{\mathcal{E}}_\text{XUV}(\omega,\tau)=\frac{\alpha}{\sqrt{2\pi}}e^{-i\omega\tau}$. Insertion into Eq.~\eqref{theory:A:response_1} yields the following expression for the response function:
\begin{widetext}
\begin{equation}
\begin{aligned}
\tilde{S}_1(\omega,\tau)&=\frac{\sqrt{\pi}\alpha^2d_{01}^2d_{12}^2\mathcal{E}^2_{0,\text{IR}}T_\text{IR}n}{4\sqrt{2}(E_1-E_2)c}\omega\cos{[\tau(E_1-\omega)]}\bigg[\frac{1}{\omega-E_1}-\frac{1}{2(E_1-E_2)} \bigg]\bigg\{-2\exp{\left[\frac{-T_\text{IR}^2(\omega-E_1)^2}{32}\right]}\\
&\quad+\exp{\left[\frac{-T_\text{IR}^2(\omega-E_1+2\omega_{\text{IR}})^2}{32}\right]}+\exp{\left[\frac{-T_\text{IR}^2(\omega-E_1-2\omega_{\text{IR}})^2}{32}\right]}\bigg\}.
\end{aligned}\label{theory:C:final_S_SB}
\end{equation}
\end{widetext}
We note that the divergence at $\omega=E_1$ can be eliminated by assigning a finite radiative lifetime, $\Gamma$, to the $E_1$ level by the substitution $E_1\to E_1-i\Gamma/2$. 

Equation~\eqref{theory:C:final_S_SB} describes the SBs and fringes, and from it we can identify the origin of these features. From Eqs.~\eqref{theory:C:d_SB1}-\eqref{theory:C:d_SB3} we see that there are two interfering sources; one originates in the field-induced time dependence of the adiabatic phases $e^{i\int_\tau^t E_{1a}(t)}$ and leads to the $1/(\omega-E_1)$ term; the other originates in the term related to the mixing of states, the complex conjugated overlap $\braket{\phi_1|\phi_{1a}(t)}^*$ between the first field-free state and the first adiabatic state, and gives rise to the constant term $-1/[2(E_1-E_2)]$. These two terms interfere constructively for $\omega<E_1$ and destructively for $\omega>E_1$, resulting in the asymmetry in the signal strength of the fringes (see Fig.~\ref{fig:fringes}). The exponentials  in Eq.~\eqref{theory:C:final_S_SB} cause Gaussian modulation about the energies $\omega=E_1$ (SBs) and $\omega=E_1\pm 2\omega_\text{IR}$ (fringes). They emerge in the Fourier transform of the squared IR field [see Eqs.~\eqref{app:A:involved} and \eqref{app:A:result}], and we note specifically that the $\exp{\left[-T_\text{IR}^2(\omega-E_1\pm2\omega_\text{IR})/32\right]}$ terms are due to the carrier factor $\sin{(\omega_\text{IR}t)}$ of the IR field. The implication is that in the presence of an IR field consisting only of an envelope part the fringes would be absent but the SBs would be present. The hyperbolic shape of the SBs and the oscillation of the fringes are both manifestations of the same factor, $\cos{\left[\tau(\omega-E_1)\right]}$, causing the features to follow hyperbolic trajectories where $\tau(\omega-E_1)=\text{constant}$~\cite{wu2016}. Thus, the present formalism presents a unified description of the SBs and the fringes, two ATAS features previously attributed to separate processes, namely perturbed free-induction decay \cite{lindberg1988,britocruz1988} and optical interference \cite{chini2014}. The cosine term comes from the $\tau$-dependent part of the phase factor, $e^{-iE_1\tau}$, combined with the $e^{-i\omega\tau}$ factor of the XUV field. Finally, we recall that Eq.~\eqref{theory:C:final_S_SB}, and by extension the SBs and fringes, came from the first term in Eq.~\eqref{theory:C:dipmom}, which is associated with the first adiabatic states, which in turn was borne out of the field-free $\ket{1s2p}$ bright state. 

\subsubsection{\label{subsubsection:LIS}Closed analytical form of LISs and discussion of physical origins} 

We return to Eq.~\eqref{theory:C:dipmom} and consider now the second term. We expand the exponential factor in orders of the IR field, insert the overlap from Eq.~\eqref{theory:C:overlap2}, and drop terms of third order or higher in the IR field. Following the same reasoning as in Sec.~\ref{subsubsection:sidebands}, we exclude $W(t-\tau)$ and set $\theta(t-\tau)=1$, yielding
\begin{equation}
\braket{d(t)}_2=-\frac{i\alpha d_{01}^2d_{12}^2}{(E_1-E_2)^2}\mathcal
{E}_\text{IR}(\tau)e^{-iE_2\tau}e^{iE_2t}\mathcal{E}_\text{IR}(t),\label{theory:C:d_LIS}
\end{equation}
from which we can obtain the frequency-dependent dipole moment (see the Appendix for details). The response function is
\begin{widetext}
\begin{equation}
\begin{aligned}
\tilde{S}_2(\omega,\tau)&=\frac{\sqrt{\pi}\alpha^2 d_{01}^2d_{12}^2\mathcal{E}^2_{0,\text{IR}}T_\text{IR}n}{2\sqrt{2}(E_1-E_2)^2c}\omega \exp{\left(\frac{-4\tau^2}{T_\text{IR}^2}\right)}\sin{\left(\omega_\text{IR}\tau\right)} \sin{\left[\tau(E_2-\omega) \right]}\times\bigg\{\exp{\left[-\frac{T_\text{IR}^2(\omega-E_2-\omega_\text{IR})^2}{16}\right]}\\
&\quad-\exp{\left[-\frac{T_\text{IR}^2(\omega-E_2+\omega_\text{IR})^2}{16}\right]}\bigg\}.
\end{aligned}\label{theory:C:final_S_LIS}
\end{equation}
\end{widetext}
Equation~\eqref{theory:C:final_S_LIS} describes the LISs, implying that their origin can be deduced from it. Unlike the case for the SBs and fringes [see Eq.~\eqref{theory:C:final_S_SB}], the LISs cannot be ascribed to the adiabatic phases. They are instead a result of the complex conjugated coefficient $a_{2}^*(t)$ corresponding to the adiabatic $\ket{\phi_{2a}(t)}$ state and the complex conjugate of its overlap with the first field-free state, $\braket{\phi_1|\phi_{2a}(t)}$ [see Eq.~\eqref{theory:C:dipmom}]. Considering first the coefficient $a_2(t)$, from Eqs.~\eqref{theory:C:adot2} we see clearly the two-photon character of the process responsible for the LISs, in agreement with previous interpretations \cite{chen2012}. Since the XUV field is approximated by a constant times a $\delta$ function [see Eqs.~\eqref{theory:C:delta} and \eqref{theory:C:a2_delta}], the subsequent integration picks out the instantaneous value of the IR field at the time $\tau$, implying that the signal strength of the LIS will follow the IR field and vanish when the field goes to zero (see Fig.~\ref{fig:LIS}). This behavior is described by the $\exp{\left(-4\tau^2/T_\text{IR}^2\right)}\sin{\left(\omega_\text{IR}\tau\right)}$ part of Eq.~\eqref{theory:C:final_S_LIS}. The overlap $\braket{\phi_1|\phi_{2a}(t)}^*$ is linear in the IR field, which upon Fourier transform leads to the terms $\exp{\left[-T_\text{IR}^2(\omega-E_2\pm\omega_\text{IR})^2/16\right]}$ that cause Gaussian modulation of the features about energies $E_2\pm\omega_\text{IR}$. Furthermore, the delay-dependent part of the phase, $e^{-iE_2\tau}$, combines with the $e^{-i\omega\tau}$ factor of the XUV field to produce the $\sin{\left[\tau(E_2-\omega)\right]}$ factor, causing the LISs to take on hyperbolic shapes, like the SBs and fringes. We note finally that the LISs originate in the second term of Eq.~\eqref{theory:C:dipmom}, corresponding to the second adiabatic state, which in turn evolved from the field-free $\ket{1s2s}$ dark state. 

\section{\label{section:results}Results}

In this section, we present the results obtained from the models described in Sec.~\ref{section:theory}. A comparison between the three-level TDSE solution and the full adiabatic three-level model is shown, and the features of main interest in the present work are highlighted. The focus then shifts to the features individually, specifically how they each change as we go from the TDSE solution to the analytic solution via the full adiabatic model.

The parameters must fulfill two criteria. We require that the fields have relatively weak intensities and the wavelength of the IR field must be chosen so that it is not in resonance with the bright-state to dark-state transition. The first condition is necessary for the validity of first-order perturbation theory out of the ground state [see Eqs.~\eqref{theory:C:adot1} and \eqref{theory:C:adot2}], and the latter condition is necessary for an adiabatic treatment of the excited states. The relatively weak intensity of the IR pulse ensures that the expansions in orders of $\mathcal{E}_\text{IR}(t)$ in Sec.~\ref{subsection:3level} are accurate. Finally, the IR wavelength must be long enough not to induce significant coupling between the three lowest levels of He and the higher levels in the bound-state manifold, to justify the use of a model with only these three levels. All generic features of interest will still be present in ATA spectra subject to these conditions, and the findings therefore also apply to other atoms where similar states and couplings can be realized.

\begin{figure}
\includegraphics[width=0.48\textwidth]{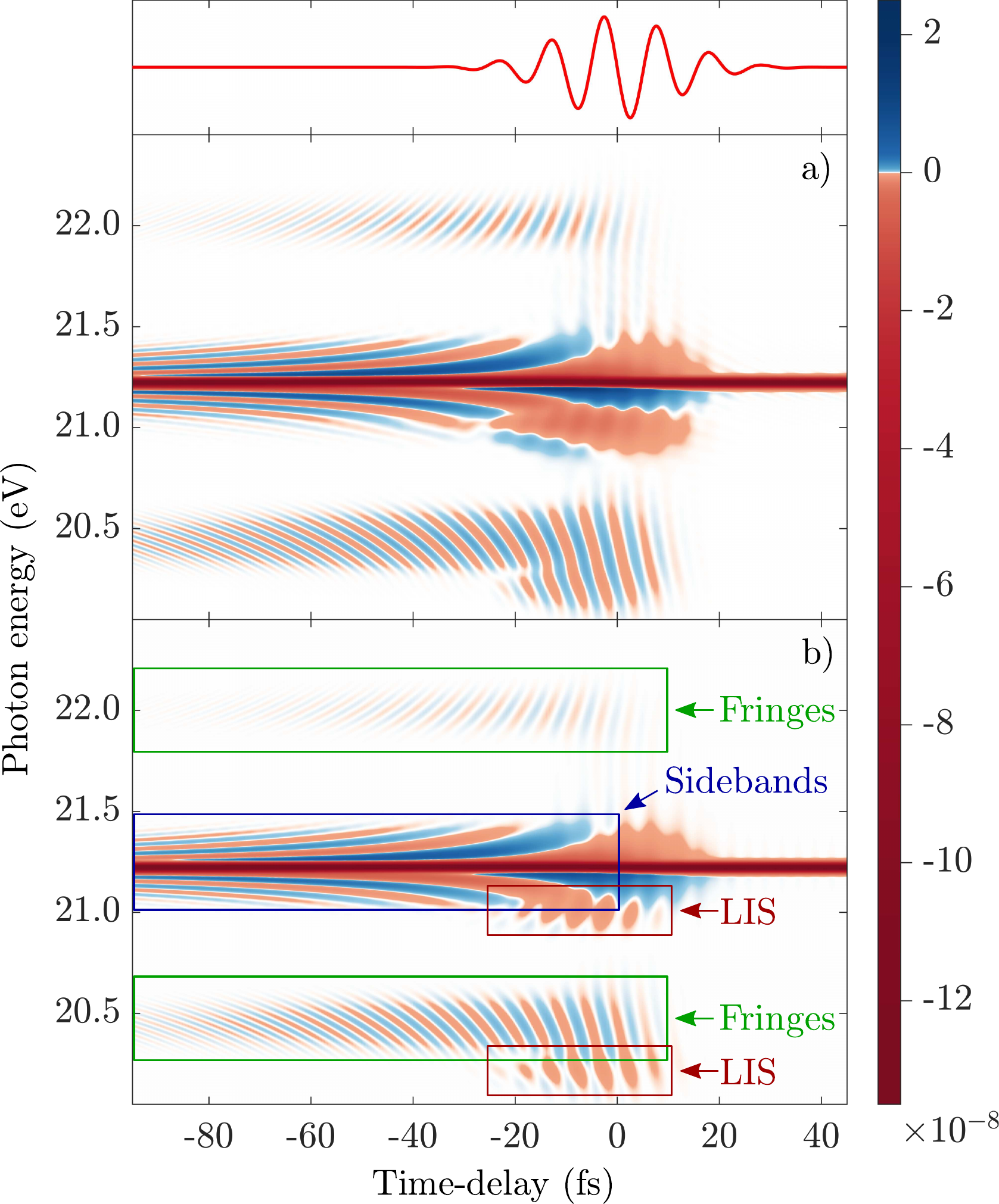}
\caption{\label{fig:TDSE3vsFULL} Attosecond transient absorption spectra $\tilde{S}(\omega,\tau)$ of He calculated by Eq.~\eqref{theory:A:response_1}. In panel (a) $\tilde{d}(\omega,\tau)$ is determined via the time-dependent Schr\"odinger equation [see Eq.~\eqref{theory:B:d_TDSE}] with a basis of $N+1=3$ levels, as in Fig.~\ref{fig:TDSE20vsTDSE3} (b). In panel (b) $\tilde{d}(\omega,\tau)$ is calculated from the full adiabatic three-level model [see Eq.~\eqref{theory:C:dipmom}] and several features of the spectrum are highlighted. Oscillating fringes are centered around $E(1s2p)\pm2\omega_\text{IR}$, i.e., at $22.00$ and $20.45$ eV; hyperbolic sidebands are centered around $E(1s2p)=21.22$ eV; and light-induced structures (LISs) are centered around $E(1s2s)\pm\omega_\text{IR}$, i.e., at $21.01$ and $20.23$ eV. The top panel illustrates the infrared pulse centered at $\tau=0$ fs. The color scale to the right shows the signal strength in arbitrary units. For pulse parameters, see the text following Eq.~\eqref{theory:A:pulse}.}
\end{figure}

In Fig.~\ref{fig:TDSE3vsFULL}, we compare the three-level TDSE results obtained from Sec.~\ref{subsection:TDSE} [see Eq.~\eqref{theory:B:d_TDSE} with $N+1=3$; Fig.~\ref{fig:TDSE3vsFULL} (a)] with the full adiabatic three-level model results obtained from Eqs.~\eqref{theory:C:eig1}, \eqref{theory:C:eig2}, \eqref{theory:C:overlap1}--\eqref{theory:C:adot2}, \eqref{theory:C:dipmom} [Fig.~\ref{fig:TDSE3vsFULL} (b)] and highlight the specific features of the spectrum that are included in the present work (as in Fig.~\ref{fig:highlights}). These features will be treated in detail individually, but are briefly described here. The features labeled ``sidebands'' are the SBs along the main absorption band of the $\ket{1s2p}$ states, which are often attributed to perturbed free-induction decay~\citep{britocruz1988,lindberg1988}. The figure shows that the SBs are very well reproduced by the adiabatic model. The features labeled ``fringes'' are oscillating with a frequency twice that of the IR field frequency and are found centered at energies $E(1s2p)\pm2\omega_\text{IR}$.  They are reproduced quite well in the adiabatic model, the main difference being a weakening of the fringes centered at $E(1s2p)+2\omega_\text{IR}$ relative to the TDSE calculations. Note that the fringes are not ``which-way'' fringes, which require additional bound states \cite{chen2012}, but rather the  type of fringes attributed to optical interference in Ref.~\cite{chini2014}. Finally, the LISs appear only in the region where there is temporal overlap between the XUV and IR field~\cite{chen2012}. They are centered at energies $E(1s2s)\pm\omega_\text{IR}$ and, like the fringes, they exhibit modulation at twice the IR frequency. There are differences between the TDSE and the adiabatic model calculations when it comes to the LISs, but characteristics such as oscillation frequency, strength, and position of the features are reproduced. 

The analytical expressions obtained in Sec.~\ref{section:theory} contain the three different classes of features shown in Fig.~\ref{fig:TDSE3vsFULL}. To support the validity of the expressions, we compare the features as they are calculated by the successive methods, first from the three-level TDSE, followed by the full adiabatic three-level model, and finally the analytical expressions. 

\begin{figure}
\includegraphics[width=0.48\textwidth]{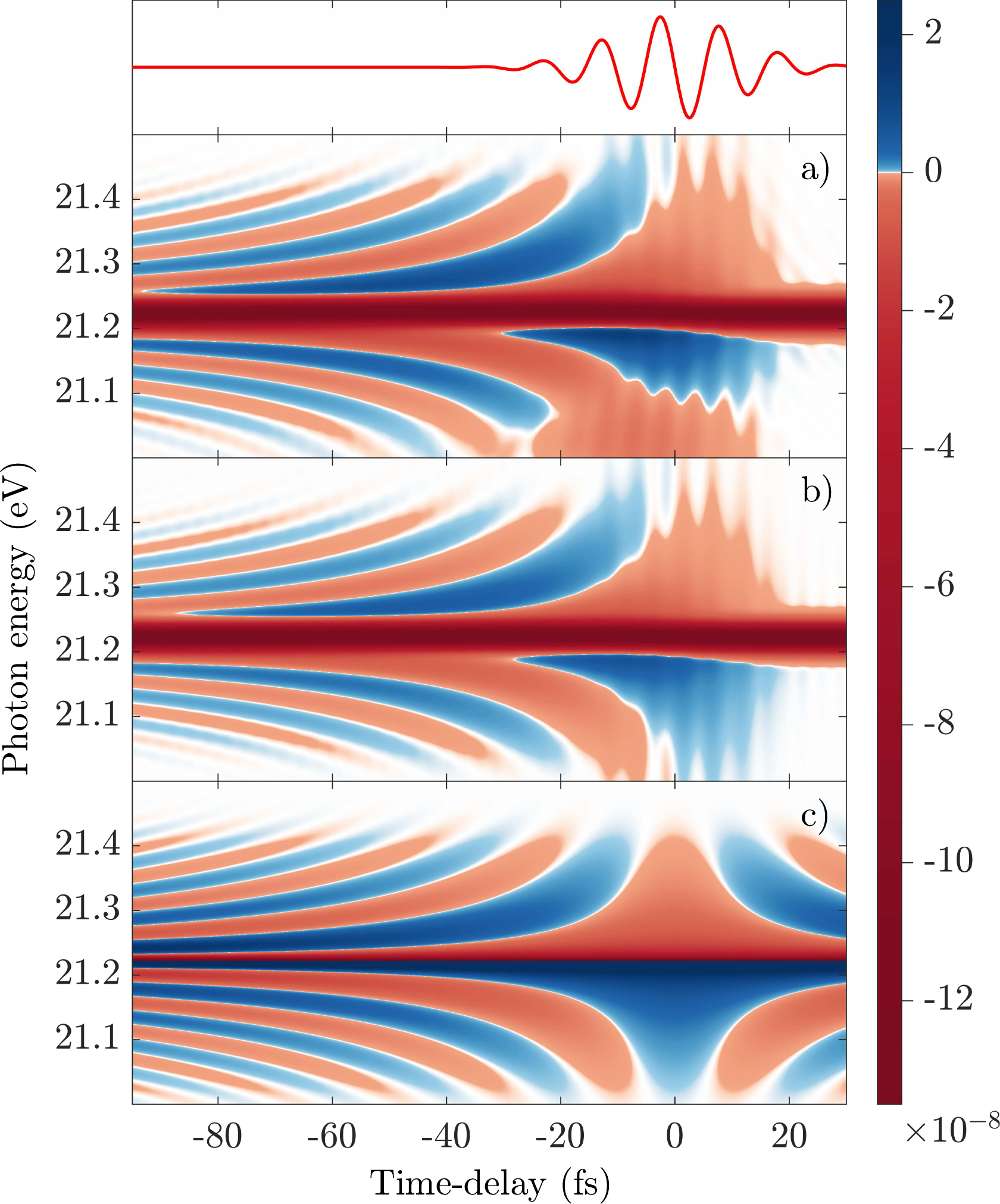}
\caption{\label{fig:SB} Attosecond transient absorption spectra $\tilde{S}(\omega,\tau)$ [Eq.~\eqref{theory:A:response_1}] focused on the hyperbolic sidebands (SBs) around the $\ket{1s2p}$ energy level of He. In panel (a) $\tilde{d}(\omega,\tau)$ is determined via the time-dependent Schr\"odinger equation [see Eq.~\eqref{theory:B:d_TDSE}] with a basis of $N+1=3$ levels, as in Fig.~\ref{fig:TDSE20vsTDSE3} (b). In panel (b) $\tilde{d}(\omega,\tau)$ is calculated from the adiabatic three-level model, with the light-induced structures seen in panel (a) (between $-25$~fs $<\tau<15$~fs, centered at $E=21.00$~eV) removed by keeping only  the first term of Eq.~\eqref{theory:C:dipmom}, resulting in an unobstructed view of the SBs. Panel (c) shows the analytic solution from Eq.~\eqref{theory:C:final_S_SB}. The top panel illustrates the infrared pulse centered at $\tau=0$ fs.  The color scale to the right shows the signal strength in arbitrary units. For pulse parameters, see the text following Eq.~\eqref{theory:A:pulse}.}
\end{figure} 

The SBs as calculated by the various methods are shown in Fig.~\ref{fig:SB}. The three-level TDSE solution can be seen in Fig.~\ref{fig:SB} (a). The adiabatic model calculations in Fig.~\ref{fig:SB} (b) are in good agreement with the TDSE results. The full adiabatic model [Eq.~\eqref{theory:C:dipmom}] consists of two terms, only the second of which is responsible for the LISs. Since, for now, we are concerned with the SBs, this second term is dropped. The same simplification cannot be attained for the numerical TDSE solution, as there is no equivalent single term accountable for the LISs. This explains the presence of a LIS (centered at $E=21.00$ eV and between $\tau=-25$ fs and $\tau=15$ fs) in Fig.~\ref{fig:SB} (a), and its absence in Fig.~\ref{fig:SB} (b). In Fig.~\ref{fig:SB} (c), the analytical solution of Eq.~\eqref{theory:C:final_S_SB} is shown, from which the term containing the $-2\exp{\left[-T_\text{IR}^2(\omega-E_1)^2/32\right]}$ factor corresponds to the SBs. The SBs are well reproduced by the analytic expressions. The differences between the analytical solution and the full adiabatic model can be attributed to the removal of separate interfering features and the suppression of $W(t-\tau)$ and $\theta(t-\tau)$ as detailed in Sec.~\ref{subsubsection:sidebands} leading up to Eq.~\eqref{theory:C:final_S_SB}.

\begin{figure}
\includegraphics[width=0.48\textwidth]{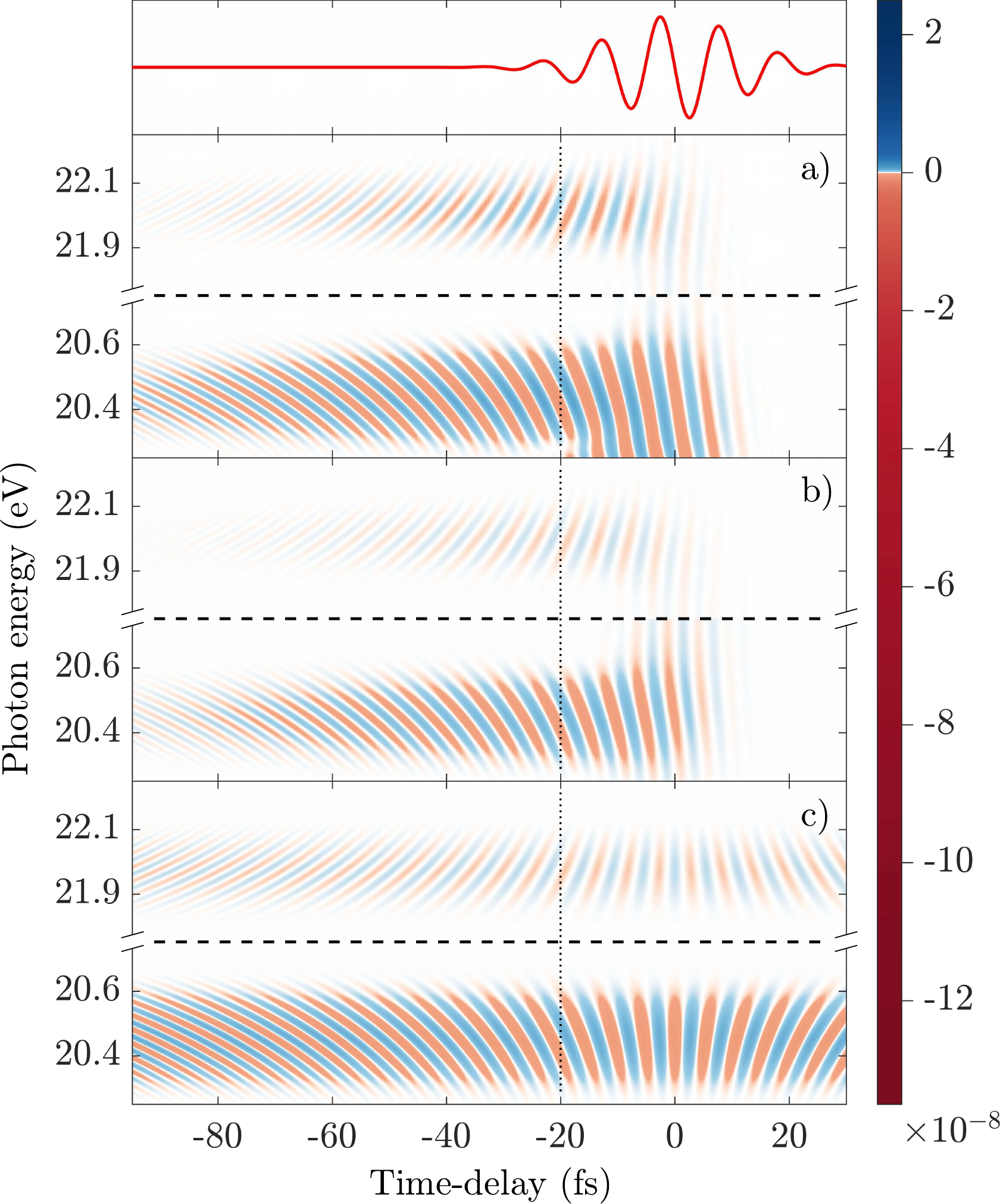}
\caption{\label{fig:fringes} Attosecond transient absorption spectra $\tilde{S}(\omega,\tau)$ [Eq.~\eqref{theory:A:response_1}] focused on the rapidly oscillating fringes around the energies $E(1s2p)\pm2\omega_\text{IR}$ in He. In panel (a) $\tilde{d}(\omega,\tau)$ is determined via the time-dependent Schr\"odinger equation [see Eq.~\eqref{theory:B:d_TDSE}] with a basis of $N+1=3$ levels, as in Fig.~\ref{fig:TDSE20vsTDSE3} (b). In panel (b) $\tilde{d}(\omega,\tau)$ is calculated from the adiabatic three-level model, with the LIS seen in panel (a) (between $-25$~fs $<\tau<15$~fs, centered at $E=20.23$~eV) removed by keeping only  the first term of Eq.~\eqref{theory:C:dipmom}; resulting in an unobstructed view of the fringes. In panel (c) the analytic solution from Eq.~\eqref{theory:C:final_S_SB} is shown. The vertical dotted line serves as a guide for the eye. The top panel illustrates the infrared pulse centered at $\tau=0$ fs. The color scale to the right shows the signal strength in arbitrary units. For pulse parameters, see the text following Eq.~\eqref{theory:A:pulse}.}
\end{figure} 

The oscillating fringes, as calculated by the various methods, are seen in Fig.~\ref{fig:fringes}. The panels above the dashed lines correspond to the fringes centered at the energy $E(1s2p)+2\omega_\text{IR}=22.00$~eV and the panels below correspond to the fringes centered at $E(1s2p)-2\omega_\text{IR}=20.45$~eV. The three-level TDSE solution can be seen in Fig.~\ref{fig:fringes} (a), and as in Fig.~\ref{fig:SB} there is a LIS (centered at $E=20.23$~eV, between $-25$~fs $<\tau<15$~fs) which should not be considered part of the comparison. We note an asymmetry in the strength of the fringes, where the fringes at $22.00$~eV are weaker; as understood from Eq.~\eqref{theory:C:final_S_SB}, this is due to interference between a term corresponding to the adiabatic phase and a term describing the mixing of the field-free states. In Fig.~\ref{fig:fringes} (b), the results from the three-level adiabatic model are shown, and except from being slightly fainter, they are in good agreement with the TDSE results. The analytic solution from Eq.~\eqref{theory:C:final_S_SB} is shown in Fig.~\ref{fig:fringes} (c), where the contribution from the term containing the $\exp{\left[-T_\text{IR}^2(\omega-E_1\pm2\omega_\text{IR})^2/32 \right]}$ factor produces the fringes. The fringes are reproduced with good qualitative and quantitative agreement; the main characteristics with which we are concerned are not affected by the differences between the analytic solution and the full adiabatic model. The causes of these differences are the same as for the SBs considered above, and are discussed in Sec.~\ref{subsubsection:sidebands}. 

\begin{figure}[t!]
\includegraphics[width=0.48\textwidth]{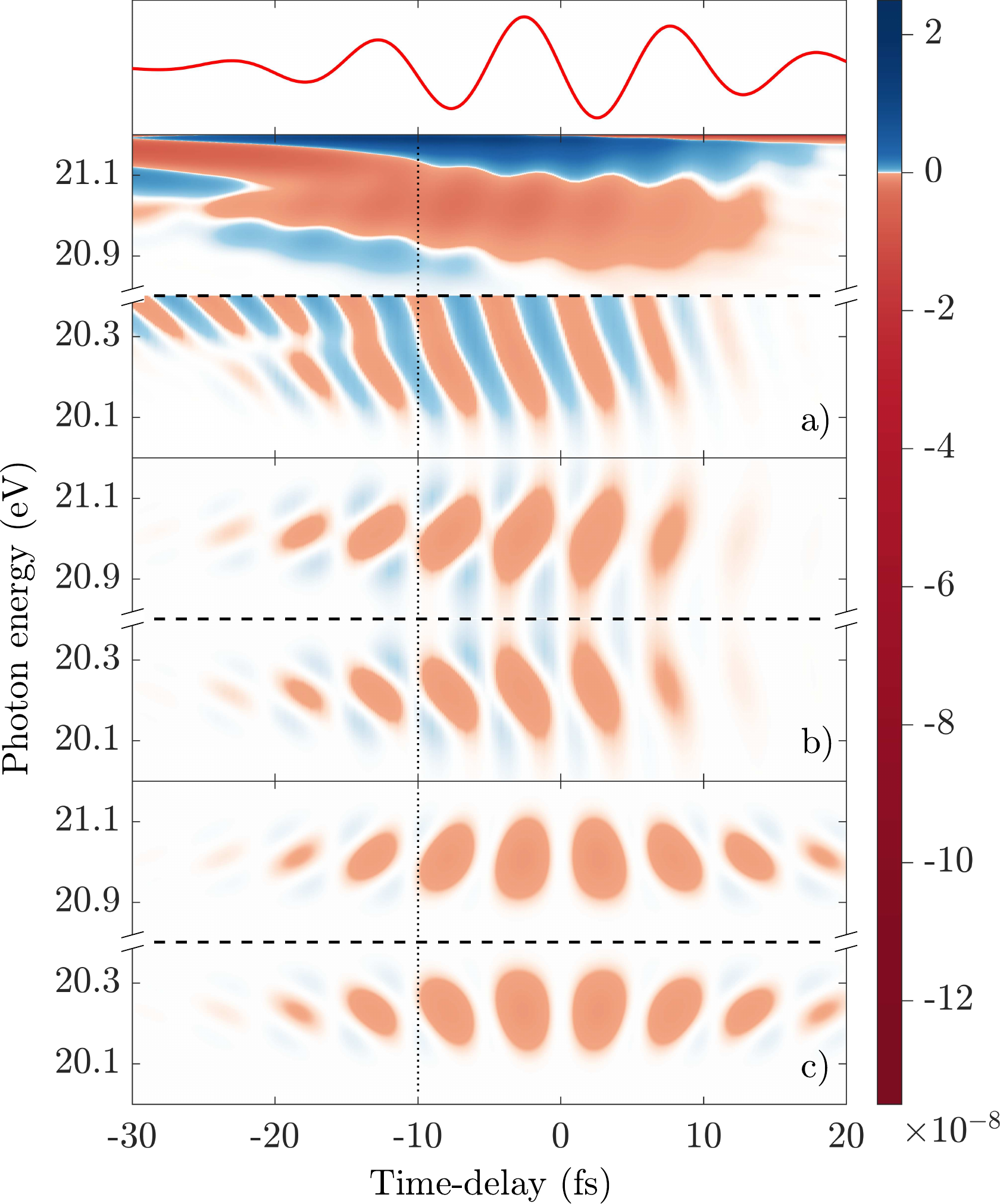}
\caption{\label{fig:LIS} Attosecond transient absorption spectra $\tilde{S}(\omega,\tau)$ [Eq.~\eqref{theory:A:response_1}] focused on the light-induced structures (LISs) around the energies $E(1s2s)\pm\omega_\text{IR}$ in He. In panel (a) $\tilde{d}(\omega,\tau)$ is determined via the time-dependent Schr\"odinger equation [see Eq.~\eqref{theory:B:d_TDSE}] with a basis of $N+1=3$ levels, as in Fig.~\ref{fig:TDSE20vsTDSE3} (b). In panel (b) $\tilde{d}(\omega,\tau)$ is calculated from the adiabatic three-level model, with the hyperbolic sidebands and oscillating fringes seen in panel (a) removed by keeping only the second term of Eq.~\eqref{theory:C:dipmom}, resulting in an unobstructed view of the LISs. Panel (c) shows the analytic solution from Eq.~\eqref{theory:C:final_S_LIS}. The vertical dotted line serves as a guide for the eye. The top panel illustrates the infrared pulse centered at $\tau=0$ fs. The color scale to the right shows the signal strength in arbitrary units. For pulse parameters, see the text following Eq.~\eqref{theory:A:pulse}.}
\end{figure} 

In Fig.~\ref{fig:LIS}, a comparison of the LISs calculated by the various methods is shown. The panels above the dashed lines correspond to the LISs centered at the energy $E(1s2s)+\omega_\text{IR}=21.01$~eV and the panels below correspond to the LISs centered at the energy $E(1s2s)-\omega_\text{IR}=20.23$~eV. The three-level TDSE solution is shown in Fig.~\ref{fig:LIS} (a), where the LISs are seen along with the other features. As explained previously, the features cannot be distinguished and separated in the TDSE solutions, something which is possible in the adiabatic model. In the top panel of Fig.~\ref{fig:LIS} (a), the LIS is mixed with the SBs and in the bottom panel the LIS is mixed with the lower oscillating fringes, somewhat complicating the comparison with the other calculations. The top LIS only takes on negative values, whereas the bottom LIS oscillates between positive and negative values similarly to the adjacent fringes. In the adiabatic model, the LISs correspond to the second term of Eq.~\eqref{theory:C:dipmom} and the other features correspond to the first term, so by keeping only the second term we can focus entirely on the LISs, the result of which is seen in Fig.~\ref{fig:LIS} (b). There is significant departure from the TDSE-based calculations in Fig.~\ref{fig:LIS} (a), but the key characteristics are still present. The features are confined to the same time delays $\tau$, the oscillation patterns match, and the strengths of the features are similar. The differences are likely attributable to the mentioned mixing of features in Fig.~\ref{fig:LIS} (a) and the approximations made in the adiabatic model calculations. The analytic solution from Eq.~\eqref{theory:C:final_S_LIS} is shown in Fig.~\ref{fig:LIS} (c), and it exhibits qualitative agreement with the full numerical solution from Fig.~\ref{fig:LIS} (b). The discrepancy beyond relatively small positive delays does not undermine the main result, as discussed for the SBs and fringes above. 

\section{\label{sec:conclusion}Conclusion and outlook}
Starting with a three-level model subject to an adiabatic condition where the excited states are the instantaneous eigenvectors of a reduced Hamiltonian, through a series of approximations we have derived closed analytic expressions describing three features commonly encountered in ATAS: fringes, SBs, and LISs [Eqs.~\eqref{theory:C:final_S_SB} and \eqref{theory:C:final_S_LIS}]. As a reference for comparison, we solved the system with a general method using the TDSE and a system corresponding to $N$ bound states of He and showed that in the relevant range of energies the spectrogram of a three-level solution agreed very well with a twenty-level solution for appropriate parameters of the fields. The three-level TDSE-based solution was compared with the full adiabatic model with agreeable qualitative and quantitative results, but with certain discrepancies due to the approximative methods implemented. Finally, the process of going from a general TDSE reference solution to analytic expressions was examined for each of the three features separately. The SBs and the oscillating fringes were well described by the analytic expression when compared to the numerical methods. Moreover, the results [Eq.~\eqref{theory:C:final_S_SB}] indicate a unification of these two features, which previously have been attributed to distinct processes, specifically perturbed free-induction decay \cite{lindberg1988,britocruz1988} and optical quantum interference \cite{chini2014}. Important characteristics of the LISs were correctly reproduced by the analytical expression. The derivation and the final form of the closed analytical expressions revealed the origin of the considered features. The expressions directly showed how the fringes, SBs and LISs depend on the properties of the pulses, such as their field strengths and the IR frequency and duration, as well as the dependence of the features on system-specific parameters, i.e., energies and transition dipole moments. Given the current experimental capabilities with respect to wavelengths and intensities, we expect that validity conditions for the model can be fulfilled in many different atomic systems.

\begin{acknowledgments}
This work was supported by the Villum Kann Rasmussen Centre of Excellence QUSCOPE. The numerical results presented in this work were
obtained at the Centre for Scientific Computing, Aarhus.
\end{acknowledgments}

\appendix*
\setcounter{equation}{0}

\section*{\label{section:appendix}Appendix: Fourier transform of $\mathbf{\braket{d(t)}_1}$ and $\mathbf{\braket{d(t)}_2}$} 

In this Appendix, we show how to obtain the Fourier transform of the time-dependent dipole moment corresponding to the SBs, fringes, and LISs in the ATA spectrum, described by Eq.~\eqref{theory:C:d_SB3}.

Following the convention 
\begin{equation}
{\mathcal{F}}[f(t)](\omega)=\frac{1}{\sqrt{2\pi}}\int_{-\infty}^\infty dt f(t) e^{-i\omega t}
\end{equation}
and starting with the dipole moment corresponding to the SBs and fringes, we wish to solve
\begin{equation}
\begin{aligned}
\tilde{d}_1(\omega,\tau)&=\frac{\alpha d_{01}^2d_{12}^2}{\sqrt{2\pi}(E_1-E_2)}e^{-iE_1\tau}\\
&\quad\times\int_{-\infty}^\infty dt \, e^{iE_1t}e^{-i\omega t}\\
&\quad\times\left[\int_\tau^tdt'\mathcal{E}_\text{IR}^2(t')+\frac{i\mathcal{E}_\text{IR}^2(t)}{2(E_1-E_2)}\right]. \label{app:A:to_solve}
\end{aligned}
\end{equation}
The factor $e^{iE_1t}$ will only cause a shift in frequency, which can be taken into consideration afterward. The field $\mathcal{E}_\text{IR}(t)$ is obtained from Eq.~\eqref{theory:A:pulse}:
\begin{equation}
\begin{aligned}
\mathcal{E}_\text{IR}(t)&=\frac{A_{0,\text{IR}}8t}{T_\text{IR}^2}e^{-\frac{4t^2}{T_\text{IR}^2}}\cos{(\omega_\text{IR} t)}\\
&\quad+\mathcal{E}_{0,\text{IR}} e^{-\frac{4t^2}{T_\text{IR}^2}}\sin{(\omega_\text{IR} t)},
\end{aligned}
\end{equation}
where we have set $t_c=0$. The IR field in our calculations has a relatively large period $T_\text{IR}$, suggesting that we can neglect the term with the factor $T_\text{IR}^{-2}$. The remaining term consists of an envelope part (the exponential factor) and a carrier part (the sine factor). 

Starting with the first of the two terms in Eq.~\eqref{app:A:to_solve}, the integral can be split up as
\begin{equation}
\int_\tau^t dt' \mathcal{E}_\text{IR}^2(t')=\int_0^tdt'\mathcal{E}_\text{IR}^2(t')+\int_\tau^0dt\, \mathcal{E}_\text{IR}^2(t), \label{app:A:split}
\end{equation}
where the second integral will be constant in time after evaluation and can be dropped, following the arguments leading up to Eq.~\eqref{theory:C:d_SB3}. Note that a splitting of the integral, as in Eq.~\eqref{app:A:split}, in the exponent on the left-hand side (LHS) of Eq.~\eqref{theory:C:expansion} prior to the expansion would impede further analytical progress. Since the integrand is even, we may rewrite the expression further as
\begin{equation}
\begin{aligned}
\int_0^tdt'\mathcal{E}_\text{IR}^2(t')&=\frac{1}{2}\int_{-t}^tdt'\mathcal{E}_\text{IR}^2(t')\\
&=\frac{1}{2}\bigg[\int_{-\infty}^{-t}dt'\mathcal{E}_\text{IR}^2(t')\\
&\quad+\int_{-t}^{t}dt'\mathcal{E}_\text{IR}^2(t')\\
&\quad-\int_{t}^{\infty}dt'\mathcal{E}_\text{IR}^2(t')\bigg]\\
&=\frac{1}{2}\int_{-\infty}^{\infty}dt' \text{sgn}(t-t')\mathcal{E}_\text{IR}^2(t'), \label{app:A:rewrite}
\end{aligned}
\end{equation}
where $\text{sgn}(t)$ is the sign function. The final expression in Eq.~\eqref{app:A:rewrite} can be recognized as the convolution of the \hrule \noindent functions  $\text{sgn}(t'-t)\equiv f(t')$ and $\mathcal{E}_\text{IR}^2(t')\equiv g(t')$, defined as 
\begin{equation}
(f*g)(t)\equiv\int_{-\infty}^\infty dt' f(t-t')g(t').
\end{equation}

From the convolution theorem, we have
\begin{equation}
\mathcal{F}[(f*g)(t)](\omega)=\sqrt{2\pi}\mathcal{F}[f(t)](\omega)\mathcal{F}[g(t)](\omega), \label{app:A:convolution}
\end{equation}
reducing the problem to finding the individual Fourier transforms of $f(t)$ and $g(t)$. The former is given by \cite{kammler2007}
\begin{equation}
\mathcal{F}[\text{sgn}(t)](\omega)=\sqrt{\frac{2}{\pi}}\frac{1}{i\omega}. \label{app:A:sgn}
\end{equation}
The latter is more involved:
\begin{equation} \label{app:A:involved}
\mathcal{F}[\mathcal{E}_\text{IR}^2(t)](\omega)=\frac{\mathcal{E}^2_{0,\text{IR}}}{\sqrt{2\pi}}\int_{-\infty}^\infty dt \exp{\left(-\frac{8t^2}{T_\text{IR}^2}\right)}\sin^2{(\omega_\text{IR}t)}e^{-i\omega t}
\end{equation}
and can be solved by expressing the sine as exponentials and completing the square, resulting in 
\onecolumngrid
\begin{equation}
\mathcal{F}[\mathcal{E}_\text{IR}^2(t)](\omega)=-\frac{\mathcal{E}_{0,\text{IR}}^2T_\text{IR}}{16}\bigg[-2\exp{\left(-\frac{T_\text{IR}^2\omega^2}{32} \right)}
+\exp{\left(-\frac{T_\text{IR}^2(\omega+2\omega_\text{IR})^2}{32}\right)}
+\exp{\left(-\frac{T_\text{IR}^2(\omega-2\omega_\text{IR})^2}{32}\right)}\bigg],
\label{app:A:result}
\end{equation}
which also solves the Fourier transform of the second term in Eq.~\eqref{app:A:to_solve}.

From Eqs. \eqref{app:A:to_solve}, \eqref{app:A:rewrite}, \eqref{app:A:convolution}, \eqref{app:A:sgn}, and \eqref{app:A:result}, and by incorporating the shift in frequency due to the factor~$e^{iE_1t}$, we obtain
\begin{equation}
\begin{aligned}
\tilde{d}_1(\omega,\tau)&=\frac{i\alpha d_{01}^2d_{12}^2\mathcal{E}^2_{0,\text{IR}}T_\text{IR}}{16(E_1-E_2)}e^{-iE_1\tau}\bigg[\frac{1}{\omega-E_1}-\frac{1}{2(E_1-E_2)} \bigg]\bigg\{-2\exp{\left[\frac{-T_\text{IR}^2(\omega-E_1)^2}{32}\right]}+\exp{\left[\frac{-T_\text{IR}^2(\omega-E_1+2\omega_{\text{IR}})^2}{32}\right]}\\
&\quad+\exp{\left[\frac{-T_\text{IR}^2(\omega-E_1-2\omega_{\text{IR}})^2}{32}\right]}\bigg\}.
\end{aligned}
\end{equation}

Next, to obtain the Fourier transform of the time-dependent dipole moment corresponding to LISs, we start with the Fourier transform of Eq.~\eqref{theory:C:d_LIS}:
\begin{equation}
\tilde{d}_2(\omega,\tau)=-\frac{i\alpha d_{01}^2 d_{12}^2}{\sqrt{2\pi}(E_1-E_2)^2}\mathcal{E}_\text{IR}(\tau)e^{-iE_2\tau}
\times\int_{-\infty}^\infty dt\, e^{iE_2t}\mathcal{E}_\text{IR}(t)e^{-i\omega t},
\end{equation}
and we make the same approximation for the IR field as above. Solving the integral then amounts to taking the Fourier transform of a Gaussian and performing shifts $E_2\pm\omega_\text{IR}$ in frequency. Thus, the Fourier transformed dipole moment responsible for the LISs is
\begin{equation}
\begin{aligned}
\tilde{d}_2(\omega,\tau)=&-\frac{\alpha d_{01}^2 d_{12}^2\mathcal{E}^2_{0,\text{IR}}T_\text{IR}}{4\sqrt{2\pi}(E_1-E_2)^2}\exp{\left(-iE_2\tau\right)}\exp{\left[\frac{-4\tau^2}{T_\text{IR}^2}\right]}\sin{\left(\omega_\text{IR}\tau\right)}\bigg\{\exp{\left[-\frac{T_\text{IR}^2(\omega-E_2-\omega_\text{IR})^2}{16}\right]}\\
&- \exp{\left[-\frac{T_\text{IR}^2(\omega-E_2+\omega_\text{IR})^2}{16}\right]}   \bigg\}.
\end{aligned}
\end{equation}
\twocolumngrid

\end{document}